\documentclass{easychair}

\usepackage{doc}     
\usepackage{cleveref}
\usepackage{tikz}
\usepackage{listings}
\usepackage{xcolor}
\usepackage{floatflt}
\usepackage{wrapfig}
\usepackage{xspace}

\lstdefinestyle{inputstyle}{
literate=
    *{:}{{{\color{blue}:}}}1
    {,}{{{\color{blue},}}}1
    {\{}{{{\color{magenta}\{}}}1
    {\}}{{{\color{magenta}\}}}}1
    {[}{{{\color{violet}[}}}1
    {]}{{{\color{violet}]}}}1
    {"players"}{{{"players"}}}9
    {"actions"}{{{"actions"}}}9
    {"infinitesimals"}{{{"infinitesimals"}}}{15}
    {"constants"}{{{"constants"}}}{11}
    {"initial_constraints"}{{{"initial\underline{ }constraints"}}}{19}
    {"property_constraints"}{{{"property\underline{ }constraints"}}}{20}
    {"weaker_immunity"}{{{"weaker\underline{ }immunity"}}}{13}
    {"weak_immunity"}{{{"weak\underline{ }immunity"}}}{15}
    {"collusion_resilience"}{{"collusion\underline{ }resilience"}}{17}
    {"practicality"}{{{"practicality"}}}{11}
    {"honest_histories"}{{"honest\underline{ }histories"}}{17}
    {"tree"}{{{"tree"}}}7
    {"player"}{{{"player"}}}{8}
    {"children"}{{{"children"}}}{10}
    {"action"}{{{"action"}}}8
    {"child"}{{{"child"}}}7
    {"utility"}{{{"utility"}}}9
    {"value"}{{{"value"}}}7
    {"A"}{{{\color{teal}"A"}}}3
    {"B"}{{{\color{teal}"B"}}}3
    {"l_A"}{{{\color{teal}"l\underline{ }A"}}}5
    {"r_A"}{{{\color{teal}"r\underline{ }A"}}}5
    {"l_B"}{{{\color{teal}"l\underline{ }B"}}}5
    {"r_B"}{{{\color{teal}"r\underline{ }B"}}}5
    {"a"}{{{\color{teal}"a"}}}3
    {"b"}{{{\color{teal}"b"}}}3
    {"a>0"}{{{\color{teal}"a > 0"}}}7
    {"a-1"}{{{\color{teal}"a-1"}}}5
    {"a-2"}{{{\color{teal}"a-2"}}}5,
  numbers=left,
  numberstyle=\scriptsize,
  stepnumber=1,
  numbersep=8pt,
  breaklines=true,
  backgroundcolor=\color{white},
  belowcaptionskip=1\baselineskip,
  frame=lines,
  xleftmargin=0.8cm,
  framexleftmargin=8mm,
  showstringspaces=false,
  basicstyle=\footnotesize\ttfamily,
  commentstyle=\itshape\color{gray},
  keywordstyle=\bfseries\color{blue},
  stringstyle=\color{orange}
}

\lstdefinestyle{outputstyle}{
  numbers=left,
  numberstyle=\scriptsize\color{black},
  stepnumber=1,
  numbersep=8pt,
  breaklines=true,
  backgroundcolor=\color{white},
  belowcaptionskip=1\baselineskip,
  frame=lines,
  xleftmargin=0.8cm,
  framexleftmargin=8mm,
  showstringspaces=false,
  basicstyle=\footnotesize\ttfamily\color{black},
  commentstyle=\itshape\color{gray},
  keywordstyle=\bfseries\color{blue},
  stringstyle=\color{orange}
}

\newcommand{\CheckMate}{\textsc{CheckMate}\xspace}
\newcommand{\A}{\texttt{A}}
\newcommand{\B}{\texttt{B}}
\newcommand{\lB}{\texttt{l\underline{ }B}}
\newcommand{\rB}{\texttt{r\underline{ }B}}
\newcommand{\lA}{\texttt{l\underline{ }A}} 
\newcommand{\rA}{\texttt{r\underline{ }A}}
\newcommand{\ma}{\texttt{a}}
\newcommand{\mb}{\texttt{b}}

\newcommand{\CommentedOut}[1]{}

\newcommand{\cpp}{C\nolinebreak\hspace{-.05em}\raisebox{.4ex}{\tiny\bf +}\nolinebreak\hspace{-.10em}\raisebox{.4ex}{\tiny\bf +}}

\title{Scaling \CheckMate for Game-Theoretic Security}

\author{Sophie Rain\inst{1}  \and Lea Salome Brugger\inst{2} \and Anja Petkovi\'c Komel\inst{1} \and Laura Kov\'acs\inst{1} \and Michael Rawson\inst{1}}

\institute{
TU Wien,
Vienna, Austria\\
\email{firstname.lastname@tuwien.ac.at}
\and
ETH Zurich,
Zurich, Switzerland\\
\email{leasalome.brugger@inf.ethz.ch}
}

\authorrunning{Rain, Brugger, Petkovi\'c Komel, Kov\'acs, Rawson}

\titlerunning{Scaling \CheckMate for Game-Theoretic Security}

\begin{document}

\maketitle

\begin{abstract}
We present the 
\CheckMate tool for automated verification of game-theoretic security properties, with application to blockchain protocols. 
\CheckMate applies automated reasoning techniques to determine whether a game-theoretic protocol model is game-theoretically secure, that is, Byzantine fault tolerant and incentive compatible.  
We describe \CheckMate's input format and its various components, modes, and output.
\CheckMate is evaluated on 15 benchmarks, including models of decentralized protocols, board games, and game-theoretic examples. 
\end{abstract}

\section{Introduction}
\label{sec:introduction}

Ensuring the security of decentralized protocols becomes even more critical in the context of decentralized finance.
Once deployed on the blockchain, vulnerabilities cannot be corrected and have the potential for significant monetary loss.
Various existing approaches for the analysis and verification of blockchain protocols~\cite{proverif, Certora, Holler23,  tamarin,  Otoni23, Tairi23, verisol} focus on cryptographic and algorithmic correctness or, in other words, whether it is possible to steal assets or gain secret information. 
However, \emph{economic} aspects must also be considered: whether it is possible for a group of users to profit from unintended behavior within the protocol itself, leading to vulnerabilities~\cite{AMHL}.
Algorithmic game theory~\cite{HalpernNE, GameTheoryBook} precisely captures such economic aspects.

This tool paper describes our open-source tool  \CheckMate\footnote{available at \url{https://github.com/apre-group/checkmate/tree/lpar25}} for the automation of game-theoretic protocol analysis. To the best of our knowledge, \CheckMate is the first fully automated tool that enforces game-theoretic security.
\CheckMate constructs and proves game-theoretic \emph{security properties} in the \emph{first-order theory of  
real arithmetic} while ensuring that game-theoretic security is precisely captured via Byzantine fault tolerance and incentive compatibility of the analyzed protocol. As introduced in our previous work~\cite{Brugger23}, Byzantine fault tolerance of a protocol guarantees that as long as users follow protocol instructions, they cannot be harmed, independently of how other users behave. Incentive compatibility ensures that the intended course of action is also the most profitable to the users, implying that no user has an economic incentive to deviate. We refer to this intended course of action as \emph{honest behavior}, captured by an \emph{honest history} in game theory. 

Following our previous work~\cite{Rain23}, inputs to \CheckMate are \emph{extensive form games} (EFGs). 
\CheckMate translates Byzantine fault tolerance into the EFG property \emph{weak(er) immunity}, whereas incentive compatibility is expressed in \CheckMate via the EFG properties \emph{collusion resilience} and \emph{practicality}. As such, protocol verification in \CheckMate becomes the task of proving weak(er) immunity, collusion resilience, and practicality, for which \CheckMate implements novel reasoning engines in first-order arithmetic.  

\emph{The purpose of this tool paper is to describe {what} \CheckMate  can do (\Cref{sec:structure}) and how it can be used (\Cref{sec:usage}).}
Theoretical details are covered in our previous work~\cite{Brugger23}, but we also improve the algorithmic setting here.
\CheckMate is no longer restricted to linear input constraints, 
improves case splitting over arithmetic formulas, and revises counterexamples to practicality, as well as weakest precondition generation and strengthening.
For efficiency reasons, \CheckMate has an entirely new implementation in \cpp, using about 2,800 lines of code tightly integrated with the satisfiability modulo theory (SMT) solver Z3~\cite{Z3}.
Our experimental results show the practical gains made over our previous work and also add 7 new benchmarks to the landscape of game-theoretic security analysis. Overall, we used \CheckMate to decide the security of 15 benchmarks, including five based on real-world protocols.

\section{Structure and Components}
\label{sec:structure}

\begin{figure}[t]
    \centering
    \includegraphics[width=\linewidth]{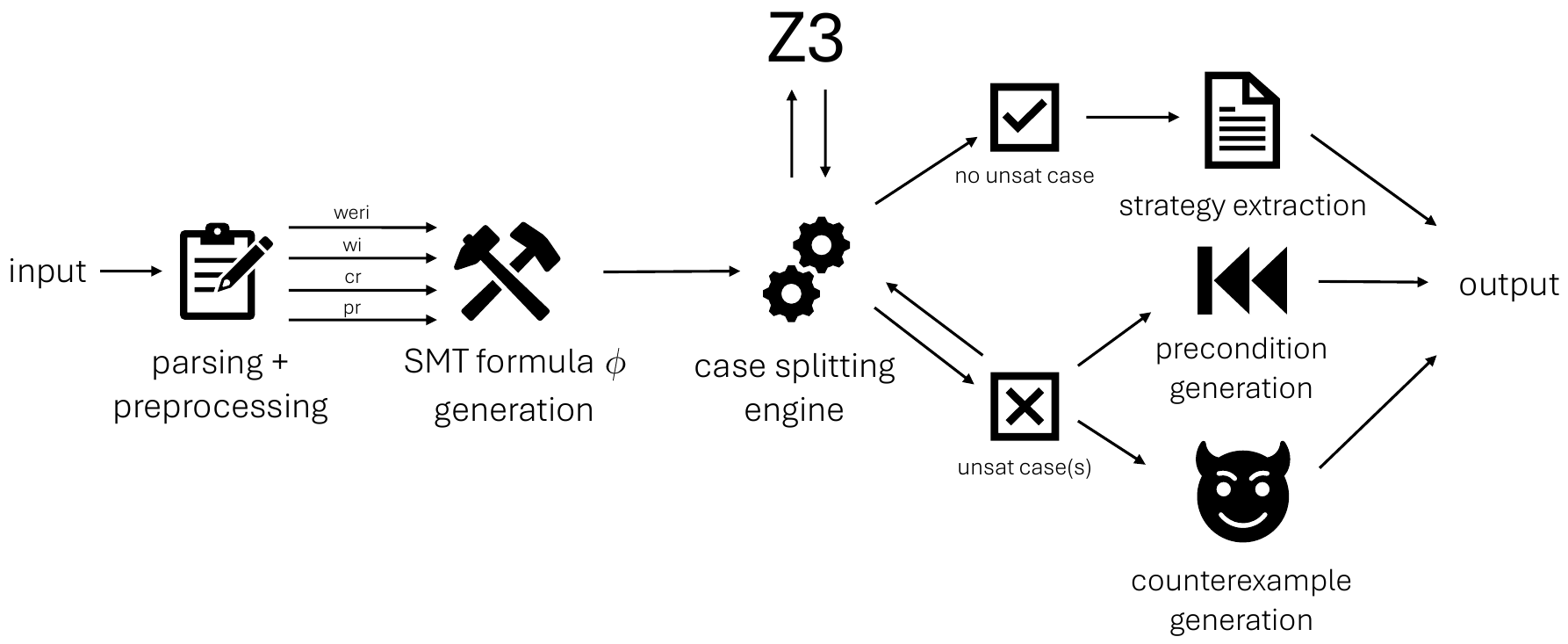}
    \caption{The \CheckMate pipeline.}
    \label{fig:CheckMate-flow-chart}
\end{figure}

\CheckMate analyzes game-theoretic security of game models. 
Given an EFG $G$, \CheckMate decides whether $G$ satisfies the security properties of (i) weak(er) immunity  -- denoted \emph{wi} respectively \emph{weri} in \Cref{fig:CheckMate-flow-chart}, (ii) collusion resilience -- \emph{cr}, and (iii) practicality -- \emph{pr}. Properties (i)-(iii) imply game-theoretic security of $G$~\cite{Brugger23}. 

\paragraph{Pipeline.} \Cref{fig:CheckMate-flow-chart} summarizes the \CheckMate pipeline.  After parsing and preprocessing an input EFG $G$, \CheckMate processes one honest history and security property (i)--(iii) at a time.
For this history and property, \CheckMate constructs \emph{an SMT formula $\phi$ such that $\phi$ is satisfiable iff the EFG satisfies the security property} -- without the need for further case analysis -- with respect to the history. 
To this end, the SMT formula $\phi$ consists of three constraints, listed as (a)--(c) in the following. The constraints  (a)--(c) are based on the central concept of encoding each action in the game as a Boolean variable that is set to true iff the corresponding action is taken in the game. In more detail, (a) 
the joint strategy constraint ensures that exactly one action is chosen at each turn. Further, (b) the honest history constraint  guarantees that the chosen set of actions yields the honest history. Finally, (c) the property constraint uses a universally-quantified formula to enforce that all variables occurring in the player's pay-offs satisfying a set of preconditions also satisfy the respective security property (i)–(iii) of the EFG.

The formula $\phi$ is passed to the case splitting engine of \CheckMate, which calls Z3 iteratively to decide whether $\phi$ is satisfiable. If not, case analysis is applied and the resulting constraints are added in turn to the preconditions of constraint (c) of $\phi$. This iterative process is terminating as \CheckMate is both sound and complete~\cite{Brugger23}. If $\phi$ is satisfiable, we extract a model -- if required by the user -- and output the result. If $\phi$ is unsatisfiable and no further case splits apply, \CheckMate implements various actions controlled by command-line options: listing cases that violate $\phi$; producing counterexamples witnessing \emph{why} $\phi$ was violated; and/or computing the weakest precondition that, if added to $G$ as an additional constraint, satisfies $\phi$. 
We describe the main components of \CheckMate using~\Cref{fig:running-example}. \newline

\begin{wrapfigure}{r}{50mm}
    \centering
    \begin{tikzpicture}
        [
            level 1/.style = {->, sibling distance = 1.6cm},
            level 2/.style = {->, sibling distance = 1.6cm}
        ]
         
        \node {$A$}
            child {
            node [yshift = 0.3cm] {$(a-1,a)$}
            edge from parent node [left, pos=0.25] {$l_A$}
            }
            child {
                node [yshift = 0.3cm]  {$B$}
                child {
                    node [yshift = 0.3cm] {$(a-2, b)$}
                    edge from parent [blue, very thick] node [left, pos=0.3] {$l_B$}
                    }
                child {
                    node [yshift = 0.3cm] {$(b, a)$}
                    edge from parent [black, line width = 0pt] node [right, pos=0.4] {$r_B$}
                    }
                edge from parent [blue, very thick] node [right, pos=0.3] {$r_A$}
                };
        \end{tikzpicture}
        \caption{Game $G$ with $a>0$, honest history {\color{blue}{$(r_A, l_B)$}}.}
    \label{fig:running-example}
\end{wrapfigure}

\noindent\textbf{Illustrative Example.}
The EFG in~\Cref{fig:running-example} has two players, $A$ and $B$.
Nodes represent the player whose turn it is and edges their choices. On reaching a leaf, the game ends, and the pay-off \emph{utility} for each player is given.
Here, player $A$ starts and chooses between actions $l_A$ and $r_A$. If  $l_A$ is chosen, the game ends, and $A$ receives utility $a-1$, whereas player $B$ receives $a$. Otherwise, $r_A$ is chosen, and player $B$  continues in a further subgame.
We assume $a>0$ and specify the  \emph{honest history} of $G$ to be 
$(r_A, l_B)$: that is, we fix the ``honest'' choice of player $A$ to be $r_A$ and of $B$ to be $l_B$.
\newline
When analyzing whether $G$ satisfies the 
 security property of weak immunity (\emph{wi}),  \CheckMate constructs the SMT formula $\phi$ with the following three
components: (a) 
    the joint strategy constraint given by $(l_A \lor r_A) \land \neg (l_A \land r_A) \land (l_B \lor r_B) \land \neg (l_B \land r_B)$; (b) 
    the honest history constraint captured by $r_A \land l_B$; and (c)
    the property constraint of $\forall a, b.\; a>0 \rightarrow \textit{wi}\,(G)$.
\subsection{\CheckMate Input}
\label{sec:input}

\CheckMate takes as input a JSON file~\cite{json} with a specific structure containing the EFG to be analyzed together with its honest histories. 
\Cref{fig:game-g-json} shows the encoding of the EFG from \Cref{fig:running-example}, conforming to the JSON schema of \CheckMate\footnote{\texttt{input.schema.json} in the repository (\url{https://github.com/apre-group/checkmate/tree/lpar25})}.
The schema defines the structure of the input as an object with the following keys:

\begin{figure}[b]
\begin{lstlisting}[style=inputstyle]
{ "players": ["A","B"],
  "actions": ["l_A","r_A","l_B","r_B"],
  "constants": ["a","b"],
  "infinitesimals": [],
  "initial_constraints": ["a>0"],
  "property_constraints": {"weak_immunity": [],
                           "weaker_immunity"  : [],
                           "collusion_resilience"  : [],
                           "practicality" : []},
  "honest_histories": [["r_A","l_B"]],
  "tree": {
    "player": "A",
    "children": [
     {"action": "l_A",
      "child": {"utility": [{"player": "A","value": "a-1"},
                            {"player": "B","value": "a"  }]}},
     {"action": "r_A",
      "child": {"player": "B",
                "children": [
                 {"action": "l_B",
                  "child": {"utility": [{"player": "A","value": "a-2"},
                                        {"player": "B","value": "b"  }]}},
                 {"action": "r_B",
                  "child": {"utility": [{"player": "A","value": "b"},
                                        {"player": "B","value": "a"}]}}]}}]}}
\end{lstlisting}
\caption{\CheckMate input encoding \Cref{fig:running-example}.}
\label{fig:game-g-json}
\end{figure}

\begin{description}
    \item[\texttt{players}] A list of all players, represented as strings.
    \item[\texttt{actions}] A list of all possible actions throughout the game, represented as strings.
    \item[\texttt{constants}] Symbolic constants occurring in the players' utilities.
    \item[\texttt{infinitesimals}] Symbolic constants occurring in the players' utilities that are treated as infinitely closer to 0 than the constants in \texttt{constants}. Symbolic values in utilities must be included either in \texttt{infinitesimals} or \texttt{constants}. 
    \item[\texttt{initial\_constraints}] Initial constraints to be enforced on the otherwise unconstrained symbolic values in utilities.
    \item[\texttt{property\_constraints}] Further initial constraints specifically for each security property of weak(er) immunity, collusion resilience,  or practicality. 
    This key lets the user specify the weakest possible assumptions for each security property.
    \item[\texttt{honest\_histories}] A list of honest histories, i.e., each history is one of the desired courses of EFG actions. Each history is the game-theoretic behavior that is (dis-)proved secure by \CheckMate sequentially. An honest history is a list of actions; therefore, this key expects a list of lists of strings. 
    \item[\texttt{tree}] The structure of the EFG. Each node in the game tree is either a branch or a leaf. Each branch is represented by an object with the following keys:
    \begin{description}
        \item[\texttt{player}] The name of the player whose turn it is.
        \item[\texttt{children}] A list of branches the player can choose from. Each branch is encoded as another object with keys \texttt{action} and \texttt{child}. The \texttt{action} key provides the action that the player takes to reach \texttt{child}, another tree.
    \end{description}
    Each leaf of \texttt{tree} is encoded as an object with a single key \texttt{utility}. As leaves represent one way of finishing the game, it contains the pay-off information for each player in this scenario. \texttt{utility} contains the players' utilities, using the following keys: 
    \begin{description}
        \item[\texttt{player}] The name of the player.
        \item[\texttt{value}] The player's utility. This can be any term over infinitesimals, constants, and reals provided as strings.
    \end{description}
\end{description}

\paragraph{\CheckMate Formulas.}
\CheckMate uses infix notation in arithmetic and Boolean expressions over real numbers, constants, and infinitesimals declared in the input.
It supports \texttt{+}, \texttt{-}, and \texttt{*} in arithmetic expressions with the usual meanings, but multiplication is allowed only if at least one of the multiplicands is not an infinitesimal.
The Boolean expressions \texttt{=}, \texttt{!=}, \texttt{<}, \texttt{<=}, \texttt{>}, and \texttt{>=} have their usual meanings.
Booleans can be combined only with disjunction spelled \texttt{|}, but this is not a limitation in practice.

\paragraph{Example (EFG in JSON format).}
In the JSON encoding of \Cref{fig:game-g-json} corresponding to the game $G$ of  \Cref{fig:running-example}, we have the following keys. The players are \A{} and \B, the actions of $G$ are \lA, \rA, \lB, and \rB. The only symbolic values of $G$ are \ma{} and \mb. None of them are supposed to be infinitesimals; thus, they are both listed under \texttt{constants}. The only initial constraint we enforce is that \ma{} is strictly positive, that is, $\texttt{a > 0}$, as specified in the caption of \Cref{fig:running-example}. We do not assume any property constraints, so the corresponding lists in \Cref{fig:game-g-json} are empty. As defined in \Cref{fig:running-example}, we consider $(\rA, \lB)$ the only honest history. 
In $G$, it is player \A{}'s turn at the first internal node, which has two children. The first child,  which is led to through action \lA{}, is an internal node containing utilities \texttt{a-1} for player \A{} and \texttt{a} for player \B{}. The other child, accessible via action \rA{}, leads to another internal node, where it is the turn of player \B{}. 

\subsection{\CheckMate Output}
\label{sec:output}
Given an input as detailed in \Cref{sec:input}, \CheckMate analyzes each specified security property (\texttt{<current property>}) for each honest history (\texttt{<current honest history>}). To this end,   \CheckMate  answers the following question in its output: 

\begin{equation}\label{CheckMate:Q}\tag{Q}
    \texttt{Is history <current honest history> <current property>?} 
\end{equation} 
 \Cref{fig:output-g-wi} shows the \CheckMate output for the input of \Cref{fig:game-g-json}, when considering the security property of weak immunity.

By answering the above question~\eqref{CheckMate:Q},  \CheckMate outputs intermediate logs about necessary case splits,  term comparisons used during splitting, and partial results during \CheckMate reasoning. Intermediate logs are displayed via indentation in the \CheckMate output (see lines 4--6 in \Cref{fig:output-g-wi}).
A partial \CheckMate result indicates the satisfiability of the considered security property in the currently analyzed case (line~6 in \Cref{fig:output-g-wi}).  
Once an answer to~\eqref{CheckMate:Q} is derived (line~8 of \Cref{fig:output-g-wi}), \CheckMate reports -- without indentation -- either
\begin{itemize}
    \item $\texttt{NO, it is not <current property>}$, in which scenario the EFG does not satisfy the analyzed security property and is, therefore, \emph{not game-theoretically secure}; 
\item $\texttt{YES, it is <current property>}$, in which case the EFG with the considered honest history has the analyzed property and \emph{may be game-theoretically secure}. If a game and a history satisfy \emph{each security property}, that is, not only the one currently analyzed but each of the three properties of weak(er) immunity, collusion resilience, and practicality, the EFG is   \emph{game-theoretically secure}. 

    \end{itemize}
\begin{figure}
\centering
\begin{lstlisting}[style=outputstyle]
    WEAK IMMUNITY
    
    Is history [r_A, l_B] weak immune?
            Require case split on (>= b 0.0)
            Require case split on (>= (- a 2.0) 0.0)
            Case [(>= b 0.0), (>= (- a 2.0) 0.0)] satisfies property.
            Case [(>= b 0.0), (< (- a 2.0) 0.0)] violates property.
    NO, it is not weak immune.
    
    Counterexample for [(>= b 0.0), (< (- a 2.0) 0.0)]:
            Player A can be harmed if:
            Player B takes action l_B after history [r_A]

    Weakest Precondition:
            (and (>= a 2.0) (>= b 0.0))
\end{lstlisting}
\caption{\CheckMate output for analyzing the weak immunity of the EFG of \Cref{fig:game-g-json}, with counterexample and weakest precondition generation.}
\label{fig:output-g-wi}
\end{figure}
In addition, 
\CheckMate can be instrumented by the user to also report on strategies (\Cref{sec:proofextraction}), 
counterexamples (\Cref{sec:counterexamples}),  and weakest preconditions (\Cref{sec:weakestpreconditions}) 
produced while answering question~\eqref{CheckMate:Q}.

\subsection{Case Splitting in \CheckMate}
\label{sec:casesplitting}
The case splitting engine takes as input the generated SMT formula $\phi$ corresponding to the analyzed security property. \CheckMate uses Z3 to determine satisfiability of $\phi$. 
If $\phi$ is satisfiable, \CheckMate uses the model satisfying $\phi$, which is provided by Z3 and proceeds to the next reasoning engine. Otherwise, Z3 reports an unsat core, a set of constraints that are a sufficient reason why $\phi$ is unsatisfiable. The case splitting engine uses this unsat core to decide whether unsatisfiability is due to (i) a necessary case split on the utilities' values that has not yet been considered; or (ii) the EFG structure.
 
If (i), \CheckMate creates two new Z3 queries: one where we add the new utility constraint to $\phi$, 
and one with its negation. The property is satisfied only if both queries are satisfiable, which might require case splitting recursively.
We record models for each case, again recursively if necessary. 
If (ii), \CheckMate records the current case split as an unsat case together with its unsat core: these are used later for counterexample and precondition generation. If requested by the user, there is also a feature to keep exploring all cases, even after encountering unsatisfiability. This allows us to provide counterexamples to unsat cases or compute weakest preconditions to be further used in redesigning protocols without unintended behavior.

\subsection{Strategy Extraction}
\label{sec:proofextraction}
If requested by the user and if the property was satisfied by the current honest history, 
\CheckMate produces explicit strategies as follows. We take as input the list of cases that we divided into in \Cref{sec:casesplitting}, together with their models,
and infer the corresponding game-theoretic strategy per case. These strategies provide a witness for the game and its honest history, satisfying the security property. The list of cases with their witness strategies is subsequently provided in the \CheckMate output.

\subsection{Counterexamples}
\label{sec:counterexamples}
If requested by the user (Section~\ref{sec:usage}) and if the honest history violated the security property, \CheckMate additionally computes counterexamples as to why the security property was violated.
Depending on further flags (\Cref{sec:usage}),  one or all counterexamples for one or all unsat cases are produced. Accordingly, the counterexamples engine receives one or all unsat cases and their unsat cores.
For all received cases, we study the unsat core to extract 
counterexamples. 

To compute all counterexamples, we forbid the game choices that led to the found counterexample by adding their negation to the SMT formula $\phi$'s constraints and checking satisfiability. We then iterate until the extended $\phi$ satisfies the property. As in previous work~\cite{Brugger23}, counterexamples to \emph{practicality} are computed differently, i.e., without the use of unsat core, while following the same iterative procedure to produce all counterexamples. 

Lines~10--12 of 
\Cref{fig:output-g-wi} list a counterexample to 
the \emph{weak immunity} of \Cref{fig:game-g-json}. Within a counterexample to weak immunity, \CheckMate reports on the harmed player $p$ (line~11 of \Cref{fig:output-g-wi}) and lists the actions the other players can take to attack $p$. These actions cannot be prevented by the honest player $p$, which leads to $p$ being harmed (line~12 of \Cref{fig:output-g-wi}).

In a counterexample to  \emph{collusion resilience}, \CheckMate provides the group of players that profits from an attack and also lists the attack. 
An attack is a set of deviating actions the malicious group takes to profit, while the honest players cannot prevent these actions. 

In a counterexample to \emph{practicality}, \CheckMate lists a player $p$, a rational subhistory $r$ different from the honest one, and the part of the honest history after which $p$ profits from deviating to $r$.

\subsection{Weakest Preconditions}
\label{sec:weakestpreconditions}

To extract the weakest precondition, that -- if added to the set of initial constraints -- makes the honest history satisfy the security property, all unsat cases have to be computed in the case splitting engine (see \Cref{sec:casesplitting}). This list of unsat cases operates as the input to the weakest preconditions routine. We apply tailored simplification steps to reduce the list of unsat cases to an equivalent and readable formula. 

This only applies if the user sets the weakest precondition flag of \CheckMate (see \Cref{sec:usage}) and the analyzed honest history violated the respective security property. In this case, the computed weakest precondition is provided as output. Lines~14--15 of \Cref{fig:output-g-wi} list the weakest precondition for the weak immunity of Figure~\ref{fig:game-g-json} and history $(r_A,l_B)$. 

\section{Usage}
\label{sec:usage}
\CheckMate invocations are of the form \texttt{checkmate GAME FLAGS}, where \texttt{GAME} is an input file as specified in \Cref{sec:input} and \texttt{FLAGS} are described below.
\CheckMate accepts the following options to modify its behavior:
\begin{description}
    \item[\texttt{--preconditions}] If a security property is not satisfied, \CheckMate computes the weakest precondition, which, if enforced additionally, would satisfy the security property. 
    \item[\texttt{--counterexamples}] If a security property is not satisfied, \CheckMate provides a counterexample showing why the property does not hold, i.e., an attack vector. 
    The number of considered scenarios is controlled by the \texttt{all\_cases} flag. 
    \item[\texttt{--all\_counterexamples}] If a security property is not satisfied, \CheckMate provides \emph{all} counterexamples for the violated case(s). 
    \item[\texttt{--all\_cases}] If a security property is not satisfied, \CheckMate computes \emph{all} violated  cases. 
    \item[\texttt{--strategies}] If a security property is satisfied, \CheckMate provides evidence in the form of a strategy that satisfies it.
\end{description}
Additionally, the user can choose which security properties to analyze with \texttt{--weak\_immunity}, \texttt{--weaker\_immunity}, \texttt{--collusion\_resilience}, and \texttt{--practicality}. If no property is specified, then all four of them will be analyzed by default.
For instance, to generate the output shown in \Cref{fig:output-g-wi}, we execute
\[
\texttt{checkmate GAME --weak\underline{ }immunity --counterexamples --preconditions} ,
\]
where {\tt GAME} is an input file containing the JSON encoding of the game in~\Cref{fig:game-g-json}.

\section{Evaluation}
\label{sec:evaluation}
We evaluated our tool on 15 benchmarks. Table~\ref{tab:results} surveys our examples, with its last 7 lines listing new benchmarks compared to~\cite{Brugger23}. 
 Out of our 15 examples, 5 describe blockchain protocols with 2, 3, or 5 players -- these are the {\bf Simplified Closing}, {\bf Simplified Routing}, {\bf Closing}, {\bf 3-Player Routing} and {\bf Unlocking Routing}. 
The {\bf Auction} example of Table~\ref{tab:results}
 models the economic behaviors of an auction;  {\bf Tic Tac Toe} as well as {\bf Tic Tac Toe Concise} a game of tic-tac-toe; whereas the 7 other examples of Table~\ref{tab:results} are game-theoretic problems with 2 to 4 players.
Table~\ref{tab:results} summarizes our experimental results. 
\CheckMate was run in default mode; that is, none of the flags described in \Cref{sec:usage} were set, and all four security properties were analyzed.
In each of the terminating benchmarks, the current \CheckMate version (version v1), presented in this paper,  is significantly faster than its initial prototype (version v0)~\cite{Brugger23}. As mentioned in \Cref{sec:introduction}, this is due to our optimized and reshaped \cpp{} implementation, as well as thanks to an improved case splitting algorithm.

\paragraph{Experimental Analysis.} Our tool improvements make even the 5-player game \textbf{Unlocking Routing}, with 36,113 nodes, feasible to analyze. 
Our biggest game \textbf{Tic Tac Toe}, on which \CheckMate does not terminate within one hour, is modeled in an unnecessarily huge way on purpose to show \CheckMate's limitations: the majority of EFG branches could be removed for symmetry reasons. A game-theoretically equivalent pruned game tree is analyzed in benchmark \textbf{Tic Tac Toe Concise}, for which \CheckMate terminates in 107.84 seconds. 
Scaling \CheckMate further is an interesting challenge for future work.

Of the 7 new benchmarks, only \textbf{Tic Tac Toe Concise} was secure for the honest history incorporating the known tie-yielding behavior. For the game-theoretic problems $G$, \textbf{Centipede}, and \textbf{EBOS}, none of the security properties hold. This is not surprising as game-theoretic problems often model dilemmas, which by their nature are not secure. Surprisingly, many of the necessary preconditions were not \texttt{false}, but rather readable and short. For example,  the weakest precondition to make the \textbf{EBOS} benchmark satisfy \emph{practicality} is $2d + f \geq p$, where $d,f$, and $p$ are variables occurring in the utilities of \textbf{EBOS}.

\textbf{Auction} violates weak immunity and collusion resilience. The model assigns the auctioneer a negative value in case the item is not being sold, while the bidders get negative values if they do not receive the item; hence,  \textbf{Auction} cannot be weak immune and is reported as such by \CheckMate. Ignoring the inconvenience of not selling the item, respectively not receiving it, \textbf{Auction} then becomes weak immune. This ignoring of small negative values is what weaker immunity incorporates \cite{Brugger23}. Further, the auctioneer and one of the bidders can collude to ensure this one bidder gets the item, which contradicts collusion resilience and is also reported by \CheckMate. The weakest precondition to imply security is \texttt{false}.

Finally, \textbf{Unlocking Routing} violates weak immunity for similar reasons as \textbf{Auction}, and thus also satisfies weaker immunity. It is practical, but is not collusion resilient, as it is vulnerable to the known Wormhole attack \cite{Rain23}. For this benchmark too, the weakest precondition to make it secure is \texttt{false}. That means the structure of the protocol has to be changed to enable security, a mere restriction of values is not enough. 

\begin{table}[!ht]
    \centering   
    \begin{tabular}{lccccc}
    \hline
        \multicolumn{1}{c}{\bf Game} & {\bf Nodes} & {\bf Players} & {\bf Histories} & {\bf Time (v1)} & {\bf Time (v0)} \\ \hline
        $\textbf{Splits}_{\textbf{wi}}$ & 5 & 2 & 3 & 0.03 & 0.35 \\ 
        $\textbf{Splits}_{\textbf{cr}}$ & 5 & 2 & 3 & 0.03 & 0.35 \\ 
        {\bf Market Entry} & 5 & 2 & 3 & 0.02 & 0.28 \\ 
        {\bf Simplified Closing } & 8 & 2 & 2 & 0.02 & 0.26 \\ 
        {\bf Simplified Routing} & 17 & 5 & 1 & 0.02 & 0.31 \\ 
        {\bf Pirate } & 52 & 4 & 40 & 1.07 & 27.08 \\ 
        {\bf Closing } & 221 & 2 & 2 &  0.34 & 9.60 \\ 
        {\bf 3-Player Routing } & 21,688 & 3 & 1 & 6.83 & 242.54 \\ 
        {\bf $G$ (\Cref{fig:running-example}) } & 5 & 2 & 1 & 0.02 & 0.18 \\ 
        {\bf Centipede } & 19 & 3 & 1 & 0.07 & 0.48 \\ 
        {\bf EBOS } & 31 & 4 & 1 & 0.02 & 0.53 \\ 
        {\bf Auction } & 92 & 4 & 1 & 0.11 & 1.72 \\ 
        {\bf Unlocking Routing } & 36,113 & 5 & 1 &  10.85 & 478.58 \\ 
        {\bf Tic Tac Toe Concise } & 58,748 & 2 & 1 & 107.84 & 254.87 \\
        {\bf Tic Tac Toe } & 549,946 & 2 & 1 & TO & TO\\
        \hline
    \end{tabular}
    \caption{Results of the current \CheckMate (v1) versus its initial prototype (v0) from~\cite{Brugger23}. Runtimes in seconds; timeout (TO) after one hour; using a 12-core AMD Ryzen 9 7900X processor running at 4.7 GHz and 128 GB of DDR5 memory clocked at 4800 MHz. 
    }
    \label{tab:results}
\end{table}
\section{Related Work and Conclusion}
\label{sec:conclusion}

We describe the \CheckMate tool for automating the security analysis of blockchain protocols.
\CheckMate complements the state of the art in protocol verification with game-theoretic security analysis, providing economic security guarantees in addition to algorithmic correctness. \CheckMate differs from existing static analyzers~\cite{Certora,Holler23,Otoni23} of Ethereum smart contracts, as these techniques merely consider cryptographic security
and formally verify the Solidity \cite{solidity} implementation of smart contracts.
Formal methods are also used in the
cryptographic verification of more general protocols~\cite{proverif,verifpal,tamarin}, yet without game-theoretic considerations.
On the other hand, existing game-based analyzers~\cite{open,prism,gambit} exhibit stochastic concurrent games and provide probabilistic results about likely behaviors~\cite{prism} or apply compositional techniques for simulating game behavior. Unlike~\cite{open,prism,gambit}, \CheckMate supports SMT-based precise reasoning over symbolic utilities without predicting/simulating its EFG properties. Security analysis in \CheckMate becomes a
theorem-proving task in first-order real arithmetic, for which \CheckMate implements novel, SMT-based techniques. With its various features and modes, \CheckMate helps blockchain developers not only to analyze their protocols but also to ``debug'' and revise their protocol modeling and verification tasks. In particular, the counterexamples generated by \CheckMate capture attack vectors to be mitigated, whereas the weakest preconditions computed by \CheckMate provide constraints to be enforced in the protocols. Our experimental results demonstrate the real-world scalability of \CheckMate, verifying, for example, the closing and routing phases of Bitcoin's Lightning Network~\cite{lightning}.

\paragraph{Acknowledgements.}  
We acknowledge funding from the ERC Consolidator Grant ARTIST 101002685, the
TU Wien SecInt Doctoral College, the FWF SFB project SpyCoDe F8504, the
WWTF ICT22-007 grant ForSmart, the Amazon Research Award 2023 QuAt, and a Netidee Fellowship 2022.

\bibliographystyle{plain}
\bibliography{references.bib}

\begin{thebibliography}{10}

\bibitem{solidity}
The~Solidity Authors.
\newblock Solidity documentation.
\newblock \url{https://docs.soliditylang.org/en/v0.8.24/}.

\bibitem{proverif}
Bruno Blanchet.
\newblock {Automatic Verification of Security Protocols in the Symbolic Model:
  The Verifier ProVerif}.
\newblock In {\em FOSAD}, pages 54--87, 2013.

\bibitem{Brugger23}
Lea~Salome Brugger, Laura Kov\'{a}cs, Anja Petkovic~Komel, Sophie Rain, and
  Michael Rawson.
\newblock {CheckMate: Automated Game-Theoretic Security Reasoning}.
\newblock In {\em CCS}, page 1407–1421, 2023.

\bibitem{Certora}
Certora.
\newblock {The Certora Prover}.
\newblock \url{https://docs.certora.com/en/latest/}.

\bibitem{Z3}
Leonardo De~Moura and Nikolaj Bj{\o}rner.
\newblock {Z3: {A}n Efficient {SMT} {S}olver}.
\newblock In {\em TACAS}, pages 337--340, Berlin, Heidelberg, 2008. Springer.

\bibitem{open}
Neil Ghani, Jules Hedges, Viktor Winschel, and Philipp Zahn.
\newblock {Compositional Game Theory}, 2016.

\bibitem{HalpernNE}
Joseph~Y. Halpern.
\newblock Beyond {N}ash {E}quilibrium: {S}olution {C}oncepts for the 21st
  {C}entury.
\newblock In {\em PODC}, page 1–10, 2008.

\bibitem{Holler23}
Sebastian Holler, Sebastian Biewer, and Clara Schneidewind.
\newblock {HoRStify: Sound Security Analysis of Smart Contracts}.
\newblock In {\em CSF}, pages 245--260, 2023.

\bibitem{json}
ISO.
\newblock The {JSON} data interchange syntax.
\newblock ISO 21778:2017, International Organization for Standardization,
  Geneva, Switzerland, November 2017.

\bibitem{verifpal}
Nadim Kobeissi, Georgio Nicolas, and Mukesh Tiwari.
\newblock {Verifpal: Cryptographic Protocol Analysis for the Real World}.
\newblock In {\em Progress in Cryptology}, pages 151--202, 2020.

\bibitem{prism}
Marta Kwiatkowska, Gethin Norman, David Parker, and Gabriel Santos.
\newblock {PRISM-games 3.0: Stochastic Game Verification with Concurrency,
  Equilibria and Time}.
\newblock In {\em CAV}, pages 475--487, 2020.

\bibitem{AMHL}
Giulio Malavolta, Pedro Moreno-Sanchez, Clara Schneidewind, Aniket Kate, and
  Matteo Maffei.
\newblock {Anonymous Multi-Hop Locks for Blockchain Scalability and
  Interoperability}.
\newblock In {\em NDSS}, 2019.

\bibitem{gambit}
Richard Mckelvey, Andrew McLennan, and Theodore Turocy.
\newblock {Gambit: Software Tools for Game Theory}, 2005.

\bibitem{tamarin}
Simon Meier, Benedikt Schmidt, Cas Cremers, and David Basin.
\newblock {The {TAMARIN} Prover for the Symbolic Analysis of Security
  Protocols}.
\newblock In {\em CAV}, pages 696--701, 2013.

\bibitem{GameTheoryBook}
Martin~J. Osborne and Ariel Rubinstein.
\newblock {\em A {C}ourse in {G}ame {T}heory}.
\newblock The MIT Press, Cambridge, USA, 1994.

\bibitem{Otoni23}
Rodrigo Otoni, Matteo Marescotti, Leonardo Alt, Patrick Eugster, Antti
  Hyv\"{a}rinen, and Natasha Sharygina.
\newblock {A Solicitous Approach to Smart Contract Verification}.
\newblock {\em ACM Trans. Priv. Secur.}, 26(2), mar 2023.

\bibitem{lightning}
Joseph Poon and Thaddeus Dryja.
\newblock The {B}itcoin {L}ightning {N}etwork: {S}calable {O}ff-{C}hain
  {I}nstant {P}ayments, 2016.
\newblock \url{https://lightning.network/lightning-network-paper.pdf}.

\bibitem{Rain23}
Sophie Rain, Georgia Avarikioti, Laura Kov{\'{a}}cs, and Matteo Maffei.
\newblock {Towards a Game-Theoretic Security Analysis of Off-Chain Protocols}.
\newblock In {\em CSF}, pages 31--46, 2023.

\bibitem{Tairi23}
Erkan Tairi, Pedro Moreno-Sanchez, and Clara Schneidewind.
\newblock {LedgerLocks: A Security Framework for Blockchain Protocols Based on
  Adaptor Signatures}.
\newblock In {\em CCS}, page 859–873, 2023.

\bibitem{verisol}
Yuepeng Wang, Shuvendu~K. Lahiri, Shuo Chen, Rong Pan, Isil Dillig, Cody Born,
  Immad Naseer, and Kostas Ferles.
\newblock {Formal Verification of Workflow Policies for Smart Contracts in
  Azure Blockchain}.
\newblock In {\em VSTTE}, pages 87--106, 2019.

\end{thebibliography}

%
\end{document}